\documentclass[twocolumn,A4]{article}
\usepackage[dvips]{graphics}
\usepackage{setspace}
\usepackage{color}
\definecolor{gold}{rgb}{0.85,0.66,0}
\definecolor{dblue}{rgb}{0,0,0.5}
\begin{document}
\onecolumn
\begin{center}
{\bf{\Large {\textcolor{gold}{Quantum transport through a conducting 
bridge: Correlation between surface disorder and bulk disorder}}}}\\
~\\
{\textcolor{dblue}{Santanu K. Maiti}}$^{1,2,*}$ \\
~\\
{\em $^1$Theoretical Condensed Matter Physics Division,
Saha Institute of Nuclear Physics, \\
1/AF, Bidhannagar, Kolkata-700 064, India \\
$^2$Department of Physics, Narasinha Dutt College,
129, Belilious Road, Howrah-711 101, India} \\
~\\
{\bf Abstract}
\end{center}
We explore a novel transport phenomenon by studying the effect of surface 
disorder on electron transport through a finite size conductor with side
coupled metallic electrodes. In the strong disorder regime the current
amplitude increases with the increase of the surface disorder strength,
while, the amplitude decreases in the weak disorder regime. This behavior
is completely opposite to that of bulk disordered system. In this article 
we also investigate the effects of the size of the conductor and the 
transverse magnetic field on electron transport and see that the transport 
properties are significantly influenced by them.
\vskip 1cm
\begin{flushleft}
\end{flushleft}
\vskip 4.5in
\noindent
{\bf ~$^*$Corresponding Author}: Santanu K. Maiti

Electronic mail: santanu.maiti@saha.ac.in
\newpage
\twocolumn

\section{Introduction}

A great challenge in the field of nanoscience and technology is to design
and fabricate electronic devices on the nanometer scale with a specific
geometry and composition. It is especially the field of molecular 
electronics that offers great potential where the electron transport is
predominantly coherent~\cite{nitzan1,nitzan2}. In $1974$, Aviram {\em et 
al.}~\cite{aviram} first studied theoretically the electron transport 
through a molecule placed between two metallic contacts, and, later several 
experiments~\cite{tali,metz,fish,reed1,reed2} have been carried out through 
different bridge systems. Electron transport through a conducting bridge 
is significantly influenced by several important factors like (a) structure 
of the conductor itself~\cite{ern2}, (b) conductor-to-electrode coupling 
strength~\cite{walc1,walc2,san1}, (c) quantum interference effect~\cite{mag,
baer1,baer2,baer3,gold,ern1} and many other important parameters that are 
needed to describe the system. These effects have been described quite 
extensively in these references and there are also lot of papers in the 
literature where we have studied different aspects of the electron 
transport, but, the complete knowledge of the conduction mechanism through 
a bridge system is still unclear even today. 

In the present paper we address a novel feature of electron transport
considering the effect of surface disorder through a finite size conductor
with side coupled electrodes. Advanced nanoscience and technology can 
easily fabricate a mesoscopic device in which charge carriers are 
scattered mainly by the surface boundaries and not by the impurities 
located in the inner core region~\cite{kou,ding1,ding2}. The idea of such 
a system named as shell-doped nanowires has been given in a recent work 
by Zhong {\em et al.}~\cite{zho} where the carrier mobility can be 
controlled nicely. The shell-doping confines the dopant atoms spatially 
within a few atomic layers in the shell region of nanowire. This is 
completely opposite to that of the conventional doping where the dopant 
atoms are distributed uniformly inside the nanowire. In this shell-doped 
nanowire system Zhong {\em et al.}~\cite{zho} have obtained a peculiar 
behavior of electron transport in which the localization becomes weaker 
with the increase of the edge disorder strength in the strong disorder 
regime, while, the localization becomes stronger in the weak disorder regime. 
This reveals that the electron dynamics in a shell-doped nanowire undergoes 
a localization to quasi-delocalization transition as the disorder increases. 
Such enhancement of the electron diffusion length or in other words the 
increment of the carrier mobility in the strong disorder regime is due to 
the existence of quasi-mobility-edges in the energy spectrum of the system. 
This finding should provide significant applications in manipulation of 
carrier transport for shell-doped nanowires of single species and as well 
as for different types of co-axial heterostructured nanowires with modulation 
doping. A similar kind of novel quantum transport in the strong disorder regime 
has also been reported in a very recent work of Zhong {\em et al.}~\cite{zho1},
where they have considered the order-disorder separated quantum films in which 
an anomalous transition is observed as the disorder strength increases. To 
reveal such an interesting phenomenon here we describe the electron transport 
through a small finite size conductor in which the impurities are located 
only in its surface boundary. From our study it is also observed that the 
electron transport through the conductor is significantly influenced by the 
size of the conductor which manifests the finite quantum size effects. 
In addition to these, here we also discuss the effect of transverse 
magnetic field on electron transport and get many interesting results. 
We utilize a simple tight-binding model to describe the system, and, 
adopt the Newns-Anderson chemisorption model~\cite{new,muj1,muj2} for 
the description of the electrodes and for the interaction of the 
electrodes to the conductor.

We organize the article as follows. In Section $2$ we present the model and 
describe the method for the calculations. Section $3$ illustrates the 
significant results, and finally, summary of the results will be
available in Section $4$.

\section{Description of model and formalism}

\begin{figure}[ht]
{\centering \resizebox*{7.5cm}{4cm}{\includegraphics{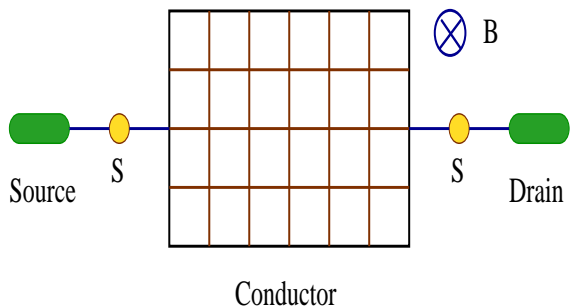}}\par}
\caption{(Color online). A finite size conductor attached to two $1$D 
semi-infinite metallic electrodes through lattice sites $S$ and $S$. 
To get a surface disordered conductor, the impurities are introduced only 
in the boundary surface which is described by the black color. While, for 
the bulk disordered case, the impurities are given in the surface as well 
as in the core regions.}
\label{bridge}
\end{figure}

This section follows the description of the model and the methodology for
the calculation of transmission probability ($T$), conductance ($g$)
and current ($I$) through a finite size conductor attached to two 
semi-infinite one-dimensional ($1$D) metallic electrodes by using the 
Green's function formalism. The schematic view of such a bridge system 
is shown in Fig.~\ref{bridge}, where the conductor is subjected to a 
transverse magnetic field $B$.

At sufficient low temperature and small applied voltage, the conductance 
$g$ of the conductor is expressed through the Landauer conductance 
formula~\cite{datta},
\begin{equation}
g=\frac{2e^2}{h} T
\label{equ1}
\end{equation}
where the transmission probability $T$ becomes~\cite{datta},
\begin{equation}
T={\mbox{Tr}} \left[\Gamma_S G_C^r \Gamma_D G_C^a\right]
\label{equ2}
\end{equation}
In this expression $G_C^r$  and $G_C^a$ are the retarded and advanced 
Green's functions of the conductor, respectively. $\Gamma_S$ and $\Gamma_D$ 
describe the couplings of the conductor to the source and drain, 
respectively. The Green's function of the conductor is written in the 
form,
\begin{equation}
G_C=\left(E-H_C-\Sigma_S-\Sigma_D\right)^{-1}
\label{equ3}
\end{equation}
where $E$ is the energy of the source electron and $H_{C}$ corresponds 
to the Hamiltonian of the conductor which can be expressed in the 
tight-binding representation within the non-interacting picture as,
\begin{equation}
H_C=\sum_i \epsilon_i c_i^{\dagger} c_i + \sum_{<ij>} t \left(c_i^{\dagger}
c_j e^{i\theta}+ c_j^{\dagger} c_i e^{-i\theta}\right)
\label{equ4}
\end{equation}
Here $\epsilon_i$'s are the on-site energies and $t$ corresponds to the
nearest-neighbor hopping strength. Here we assume that the hopping strengths
along the longitudinal and the transverse directions are identical to each
other, for simplicity. The phase factor $\theta=(2\pi/\phi_0)\int_0^a 
\vec{A}.\vec{dl}$, where $a$ is the lattice spacing and $\vec{A}$ ($By,0,0$) 
is the vector potential associated with the magnetic field $B$. In order 
to introduce the impurities in the conductor we set the site energies 
($\epsilon_i$'s) in the form of incommensurate potentials through the 
expression $\epsilon_i=\sum_i W \cos(i \lambda \pi)$ where $\lambda$ is 
an irrational number, and as a typical example, we take it as the golden mean 
$\left(1+\sqrt{5}\right)/2$. $W$ is the strength of disorder. Setting 
$\lambda=0$ we get back the pure system with identical site potential $W$. 
Now to describe the two electrodes we use a similar kind of tight-binding 
Hamiltonian as prescribed in Eq.(\ref{equ4}), where the on-site energy and 
the nearest-neighbor hopping strength are described by the parameters 
$\epsilon_i^{\prime}$ and $v$, respectively. In Eq.(\ref{equ3}), $\Sigma_S$ 
and $\Sigma_D$ represent the self-energies due to the coupling of the 
conductor to the source and drain, respectively, where all the 
information of the coupling are included into these two self-energies 
and are described through the use of Newns-Anderson chemisorption 
model~\cite{new,muj1,muj2}. In our tight-binding formulation the hopping 
strength of the conductor to the two metallic electrodes are represented 
by the parameters $\tau_S$ and $\tau_D$, respectively. By utilizing the 
Newns-Anderson type model we can express the conductance in terms of the 
effective conductor properties multiplied by the effective state densities 
involving the coupling. This permits us to study directly the conductance 
as a function of the properties of the electronic structure of the 
conductor within the electrodes.

The current passing across the conductor can be assumed as a single electron
scattering process between the two reservoirs of charge carriers. The
current-voltage ($I$-$V$) characteristics can be computed through the 
expression~\cite{datta},
\begin{equation}
I(V)=\frac{e}{\pi \hbar}\int \limits_{-\infty}^{\infty} 
\left(f_S-f_D\right) T(E)~ dE
\label{equ5}
\end{equation}
where $f_{S(D)}=f\left(E-\mu_{S(D)}\right)$ gives the Fermi distribution
function with the electrochemical potential $\mu_{S(D)}=E_F\pm eV/2$.
For the sake of simplicity, here we assume that the entire voltage is
dropped across the conductor-to-electrode interfaces and this assumption 
does not significantly affect the qualitative aspects of the current-voltage
characteristics. Such an assumption is based on the fact that the electric 
field inside the conductor, especially for short conductors, seems to have 
a minimal effect on the $g$-$E$ characteristics. On the other hand, for 
bigger conductors and higher bias voltage, the electric field inside the 
conductor may play a more significant role depending on the size and the 
structure of the conductor~\cite{tian}, yet the effect is much small.

Throughout the paper we describe our results at very low temperature
($4$ K), but the qualitative features of all the results are invariant
up to some finite temperature ($\sim 300$ K). For simplicity we take the
units $c=e=h=1$ in our present discussion.

\section{Results and discussion}

Here we focus our results and discuss the correlation effect between the 
surface disorder and bulk disorder on electron transport through a finite 
size conductor with side coupled metallic electrodes. The effects of the 
transverse magnetic field and the size of the conductor are also described 
in the subsequent parts. Our results suggest the principles for 
understanding and control of the electron transport through any bridge 
system. Here we use the parameters $N_x$ and $N_y$ to denote the total 
number of atomic sites of the conductor along the $x$ and $y$ directions, 
respectively. We choose $N_y$ as odd always and connect the electrodes 
symmetrically to the conductor as presented in Fig.~\ref{bridge}. Throughout 
our study we set the values of the different parameters as follows: the 
\begin{figure}[ht]
{\centering \resizebox*{7.5cm}{5cm}{\includegraphics{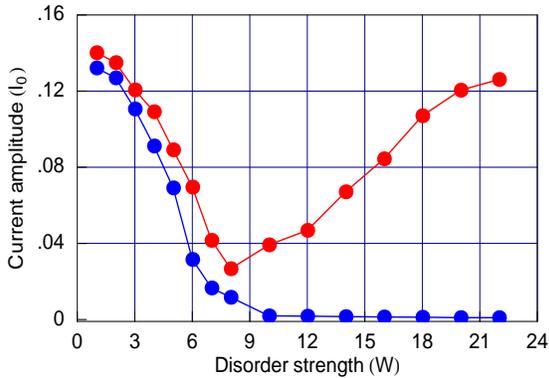}}\par}
\caption{(Color online). Typical current amplitude $I_0$ as a function 
of the disorder strength $W$ for the conductor with $N_x=6$ and $N_y=5$, 
in the absence of any magnetic field i.e., $B=0$. The red and blue curves 
correspond to the results for the surface and bulk disordered conductors, 
respectively.}
\label{current1}
\end{figure}
hopping strengths of the conductor to the electrodes (source and drain) 
$\tau_S=\tau_D=0.5$ and the nearest-neighbor hopping integral $t=3$. In the 
electrodes, we fix the hopping strength $v$ between the nearest-neighbor 
sites at $3$ and the on-site energies ($\epsilon_i^{\prime}$'s) of all 
the atomic sites in these electrodes are taken as zero. The Fermi energy 
$E_F$ is set to $0$.

In Fig.~\ref{current1} we plot the typical current amplitudes $I_0$ as a 
function of the disorder strength $W$ for a finite size conductor
considering $N_x=6$ and $N_y=5$. The transverse magnetic field $B$ is 
set at $0$ and the typical current amplitudes are computed at the applied 
bias voltage $V=1$. The red and blue curves correspond to the results for 
the surface and bulk disordered conductors, respectively. Here we 
introduce only the diagonal disorder considering the on-site energies from 
the incommensurate potential distribution function as stated earlier, and 
in the obvious reason we do not take any disorder averaging. For the bulk 
disordered conductor, the current amplitude gradually decreases with the 
increase of the impurity strength $W$. This behavior can be clearly
understood from the theory of Anderson localization where we get more
\begin{figure}[ht]
{\centering \resizebox*{7.5cm}{5cm}{\includegraphics{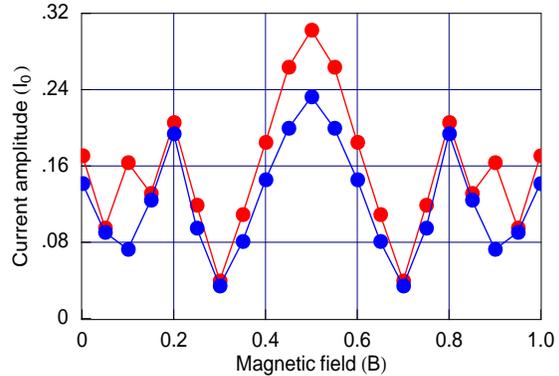}}\par}
\caption{(Color online). Typical current amplitude $I_0$ as a function 
of the magnetic field $B$ for a conductor with $N_x=6$ and $N_y=5$ in 
the weak disorder ($W=3$) regime. The red and blue curves correspond 
to the results for the surface and bulk disordered cases,respectively.}
\label{field1}
\end{figure}
localization with higher disorder strength and this is a well known feature 
in the study of electron transport. A dramatic feature is observed for the 
conductor when the impurities are introduced only in its surface boundary.
The current amplitude decreases initially with the increase of the impurity
strength and after reaching a minimum it again increases with the strength 
of the impurity. This is completely opposite in nature from the bulk 
disordered case, where current amplitude always decreases with the impurity
strength. Such an anomalous behavior for the surface disordered case
can be explained in the following way. We can treat the surface disordered 
conductor as a coupled system where the disordered surface is coupled to 
the inner core perfect region, according to the description given by Zhong 
{\em et al.}~\cite{zho}. Thus we can predict that, in the absence of any 
coupling, the localized states are obtained at the surface region, while we 
get the extended states in the inner core perfect region. For the coupled 
system, the coupling between these localized states to the inner core 
extended states is strongly influenced by the strength of surface 
disorder. In the limit of weak disorder the coupling is strong, while the 
coupling effect becomes less important in the limit of strong disorder. 
Therefore, in the weak disorder regime the electron transport is strongly 
influenced by the impurities at the surface in which the electron states 
are scattered more, and accordingly, the current amplitude decreases. On 
the other hand, for the stronger disorder regime the inner core extended 
states are less influenced by the surface disorder and the coupling effect 
\begin{figure}[ht]
{\centering \resizebox*{7.5cm}{5cm}{\includegraphics{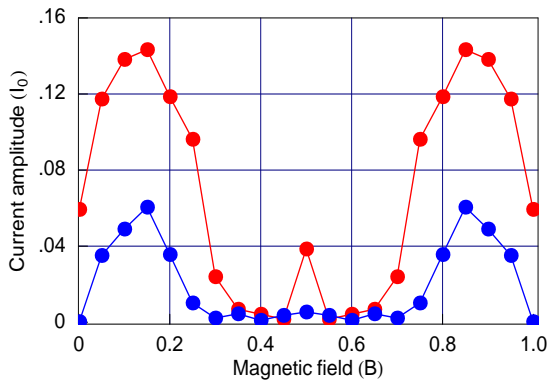}}\par}
\caption{(Color online). Typical current amplitude $I_0$ as a function 
of the magnetic field $B$ for a conductor with $N_x=6$ and $N_y=5$ in 
the strong disorder ($W=15$) regime. The red and blue curves correspond 
to the identical meaning as in Fig.~\ref{field1}.}
\label{field2}
\end{figure}
gradually decreases with the impurity strength. Therefore, the current 
amplitude increases gradually in the strong disordered regime. For the 
large enough impurity strength, the inner core extended states are almost 
unaffected by the impurities at the surface and in that situation we get 
the current only due to the inner core extended states which is the trivial 
limit. So the exciting limit is the intermediate limit of $W$.

To investigate the effect of the transverse magnetic field on electron
transport now we describe the results those are plotted in Figs.~\ref{field1}
and \ref{field2}. Here we take the same system size as considered in
Fig.~\ref{current1}. The red and blue curves correspond to the surface
and bulk disordered conductors, respectively, where Fig.~\ref{field1}
represents the behavior of the current amplitudes in the weak disorder
($W=3$) regime and the results for the limit of strong disorder ($W=15$)
are shown in Fig.~\ref{field2}. The typical current amplitudes are 
calculated at the same bias voltage as mentioned earlier. Both for 
the two limiting cases of the disorder strength, the current amplitude 
varies periodically with the magnetic field $B$, showing $\phi_0$ ($=ch/e$,
the elementary flux-quantum) periodicity. In the limit of weak disorder, 
the current amplitudes both for the surface 
and bulk disordered conductors are comparable to each other in the 
whole range of $B$ showing a maximum around $B=0.5$ (Fig.~\ref{field1}). 
On the other hand, for the case of strong disorder, the current amplitudes 
are comparable only in the small range around $B=0.5$ (Fig.~\ref{field2}), 
while for all other ranges the current amplitude in the surface disordered 
conductor is very large than that of the bulk disordered case which indicates 
that in these ranges the inner core extended states are much less affected 
by the surface disorder. All these results are also valid for the conductors 
\begin{figure}[ht]
{\centering \resizebox*{7.5cm}{5cm}{\includegraphics{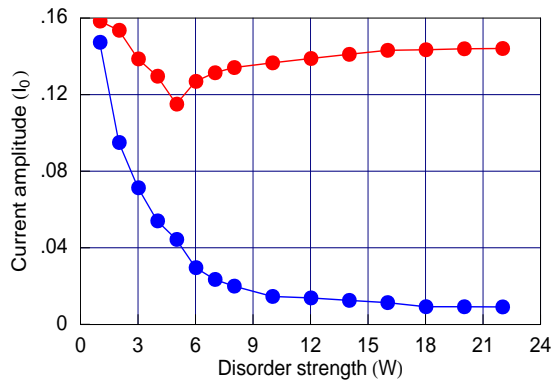}}\par}
\caption{(Color online). Typical current amplitude $I_0$ as a function of 
the disorder strength $W$ for a conductor with $N_x=7$ and $N_y=5$, in the 
absence of any magnetic field. The red and blue curves represent the 
identical meaning as in Fig.~\ref{current1}.}
\label{current2}
\end{figure}
with other system sizes ($N_x$ and $N_y$). Thus we can predict that in the 
limit of strong disorder the correlation among the surface disorder and 
bulk disorder strongly depends on the strength of the transverse magnetic 
field $B$. 

To emphasize the finite quantum size effects on electron transport, in 
Fig.~\ref{current2} we plot the results for the bridge system considering
$N_x=7$ and $N_y=5$. The results are plotted for $B=0$ those are computed at
the typical bias voltage $V=1$ and the red and blue curves represent 
the identical meaning as in Fig.~\ref{current1}. In the two disordered 
regimes the characteristic features of the current amplitudes both for the 
surface and bulk disordered conductors are similar to that as given in 
Fig.~\ref{current1}. But the significant observation is that the overall 
current amplitude for this surface disordered conductor is much larger than 
that of the results as observed for the surface disordered conductor with 
$N_x=6$ and $N_y=5$ (see the red curve of Fig.~\ref{current1}). This behavior 
can be explained as follows. For a fixed $N_y$ ($N_y=5$) as we increase the 
system size from $N_x=6$ to $N_x=7$, the ratio of the surface to inner
core region decreases ($1.5$ to $1.33$), and accordingly, the surface effect 
becomes less important. Therefore, the current carried by the inner core 
region will be less affected by the surface disorder which provides greater 
current amplitude. Now, if we change the system size in such a way that the 
ratio of the surface to inner core region increases, then we will get 
lesser current in the overall region. Another important observation is that, 
the typical current amplitude where it goes to a minimum strongly depends 
on the system size which is clearly visible from the red curves plotted in 
Figs.~\ref{current1} and \ref{current2}. These results reveal the finite
quantum size effects in the study of electron transport phenomena.

\section{Concluding remarks}

To summarize, we have studied a peculiar effect of the surface disorder 
on quantum transport through a finite size conductor with side coupled 
electrodes by using the Green's function technique, based on the 
tight-binding formulation. For a surface disordered conductor our results 
have provided an anomalous behavior in which the current amplitude 
increases with the increase of the disorder strength in the strong 
disorder regime, while the current amplitudes decreases in the weak 
disorder regime. This feature is completely opposite to that of the 
bulk disordered conductor in which the current amplitude decays gradually 
with the increase of the impurity strength. The model studied here is a 
generalization of the novel class of quantum structures named as 
shell-doped nanowires~\cite{zho}, order-disorder separated quantum 
films~\cite{zho1}, where the anomalous transition is observed in the 
strong disorder regime. From our study of the electron transport in 
the presence of transverse magnetic field we have observed that, in 
the limit of strong disorder the correlation among the surface disorder 
and bulk disorder strongly depends on the strength of the magnetic field. 
Finally, to emphasize the finite quantum size effects we have studied 
the electron transport by varying the system size and the results have 
predicted that the typical current amplitude where it goes to a minimum 
significantly depends on the size of the conductor.

Throughout our study we have considered several important assumptions by 
neglecting the effects of all the inelastic scattering processes, the 
electron-electron correlation, the Schottky effect, the static Stark 
effect, etc. More studies are expected to take into account all these 
assumptions for our future study.

\end{document}